\documentclass{article}
\usepackage{graphicx,algorithmic,hyperref,url}
\usepackage{amsmath,amssymb,amsthm}

\newcommand{\bN}{\mathbb{N}}
\newcommand{\bZ}{\mathbb{Z}}
\newcommand{\bR}{\mathbb{R}}
\newcommand{\ba}{\begin{eqnarray}}
\newcommand{\ea}{\end{eqnarray}}
\newtheorem{problem}{Problem}

\newtheorem{theorem}{Theorem}[section]
\newtheorem{corollary}[theorem]{Corollary}
\newtheorem{proposition}[theorem]{Proposition}
\newtheorem{remark}[theorem]{Remark}
\newtheorem{definition}[theorem]{Definition}
\newtheorem{example}{Example}[section]
\newtheorem{lemma}[theorem]{Lemma}
\begin{document}
\title{Optimal rounding under integer constraints}
\author{Rama Cont\footnote{Mathematical Institute, University of Oxford. E-mail: {\tt Rama.Cont@maths.ox.ac.uk}}\ \  and Massoud Heidari\footnote{LeastBit Information Lab.} }
\date{2026}
\maketitle
\begin{abstract}
Given $N$ real numbers whose sum is an integer, we study the problem of
finding $N$ integers that preserve the sum while minimizing the rounding
error. We first show that every optimal solution necessarily rounds each
coordinate either to its floor or its ceiling, reducing the problem to the
selection of the coordinates to be rounded upward. This characterization
extends to a class of separable convex integer optimization problems with a
single sum constraint.

For the resulting optimization problem we characterize the complete set of
optimal solutions and show that rounding upward the largest fractional parts
simultaneously minimizes the $L^q$ rounding error for every $q\geq 1$.
More generally, the resulting error vector is minimal in the weak-majorization
order and therefore minimizes every symmetric convex coordinatewise nondecreasing
loss of the rounding error. When the $L^q$-optimal solution is not unique, we provide an explicit
tie-breaking rule that minimizes the relative rounding error among all
optimal solutions.

These structural results lead to a deterministic algorithm with linear
$O(N)$ worst-case complexity. Unlike independent randomized rounding,
which preserves the target coordinates and the sum constraint only in
expectation, the proposed method computes an exactly feasible,
provably optimal integer rounding with deterministic optimality guarantee. Besides solving the constrained rounding
problem, the algorithm applies as the rounding step in relaxed integer
optimization problems with a single conservation constraint.
\end{abstract}

\newpage
\tableofcontents
\newpage
\section{Introduction}


The  rounding, or integer approximation, of real numbers  is a key step in many algorithms used in integer optimization \cite{grotschel1993,korte2002,lenstra1990,williamson2011}, whereby 
\begin{itemize}
\item[(i)] an optimization problem over integers is replaced by  a corresponding problem over real numbers ('relaxation' step);
\item[(ii)] the real solution obtained in (i) is 'rounded' to obtain an integer solution satisfying the constraints of the original problem (rounding step).
\end{itemize}
If the objective function  of the relaxed problem is linear, convex or has some other special property, step (i) may be solved using efficient numerical methods with polynomial complexity in the dimension $N$ of the problem.
The rounding step, on the other hand, may be described as a constrained integer optimization problem over $\{0,1\}^N$ for which  no  generic polynomial algorithm is known. 

Given  the large dimensionality of integer programming problems arising in many applications, exhaustive search on  $\{0,1\}^N$ is certainly not an option. Randomized rounding has been extensively used and analyzed as an alternative \cite{raghavan1987}. Yet in many instances randomized rounding may result in a substantial loss in accuracy \cite{williamson2011}. 
Thus, in practice, the final rounding step sets a lower bound on the overall accuracy of many integer  optimization algorithms, which illustrates the importance of good rounding algorithms.

Many integer optimization problems, such as scheduling of tasks on parallel machines \cite{potts1985,lenstra1990,schulz2002},  or the proportional allocation  of seats  to political parties in elections with party list voting systems \cite{balinski,niemeyer2008}, are naturally subject to integer constraints. The latter problem, known as the {\it apportionment problem}, has a long history in mathematics, dating back to Polya \cite{polya1919,polya1919b}, and continues to generate a lot of interest \cite{grimmett2012,janson2014}.
Relaxation to real variables followed by rounding each component to the nearest integer may typically fail to satisfy such constraints and many different --and non-equivalent-- methods exist for obtaining approximate or exact solutions satisfying the integer constraints \cite{balinski,raghavan1987,niemeyer2008,williamson2011}. Most of  these methods are based on {\it heuristics} which provide approximate solutions. Some algorithms, based on randomized rounding  may yield a bias in finite samples, which may or may not vanish asymptotically \cite{diaconis1979,janson2014}.
Even in the case where they yield exact solutions,   the mathematical properties of such algorithms have not been systematically analyzed  until  recently \cite{niemeyer2008}.

We complement this picture by demonstrating the link between various families of such {\it optimal rounding} problems, proposing an algorithm with deterministic linear complexity for solving them, giving a systematic analysis of the optimality and computational complexity of this algorithm and comparing it to some popular methods.

\subsection{Related literature}

The problem considered here is related to several strands of the literature on
integer optimization, constrained rounding, and resource allocation.

A common paradigm in mixed-integer optimization is to solve a continuous
relaxation and subsequently round the relaxed solution.
Berthold \cite{BertholdRENS} introduced the RENS heuristic, which restricts
integer variables to their floor or ceiling values around a relaxation point and
solves the resulting mixed-integer subproblem.
The problem considered here is a structured special case in which the only
constraint is preservation of the total sum and the objective is the rounding
error. In this setting we show that the floor-or-ceiling restriction is not a
heuristic but a {\it necessary} property of every optimum, allowing the subproblem
to be solved explicitly in linear time.

The common-loss separable subclass of Problem~\ref{problem:2} considered in
Proposition~\ref{prop:2} belongs to the class of separable convex resource
allocation problems.

Polynomial algorithms for such problems have been studied by
Federgruen and Groenevelt \cite{FedergruenGroenevelt1986},
Ibaraki and Katoh \cite{IbarakiKatoh1988}, and
Hochbaum and Shanthikumar \cite{HochbaumShanthikumar1990}.
Our contribution is complementary to these general results.
Exploiting the common-loss structure and the equality constraint, we obtain a
complete characterization of the optimal integer solutions, prove simultaneous
optimality for every $L^q$ norm, establish weak-majorization optimality, and
derive a deterministic $O(N)$ algorithm.

The paper also relates to the literature on randomized rounding.
Independent randomized rounding \cite{raghavan1987} preserves the target point
only in expectation and does not, in general, satisfy the sum constraint.
More sophisticated dependent rounding schemes
\cite{Gandhi2006} preserve cardinality or other linear constraints while
retaining prescribed marginals.
Unlike these randomized methods, our algorithm is deterministic and computes
an exact minimizer of the constrained rounding problem. 

The decimal-rounding problem considered in Section~\ref{sec.problems} is a single-total instance of controlled rounding, introduced by Cox and Ernst
\cite{CoxErnst1982}, while the preservation of cumulative sums studied by
Knuth's two-way rounding algorithm \cite{Knuth1995} addresses a different
rounding criterion based on discrepancy rather than optimal approximation in an
$L^q$ norm.

Finally, our primary rounding rule coincides with the classical
largest-remainder (Hare--Niemeyer) method used in proportional apportionment
\cite{balinski,Pukelsheim2014}. 
The present work provides an optimization interpretation of this rule
by showing that it minimizes every \(L^q\) rounding error and, more
generally, every symmetric, convex, coordinatewise nondecreasing
function of the absolute rounding-error vector.
The proposed relative-error tie-breaking rule further selects a relative-error minimizer among multiple $L^q$-optimal allocations.

\subsection{Outline}

We formulate the {\it optimal rounding} problem in Section~\ref{sec:optimalrounding}. In Section~\ref{sec:main}, we establish the componentwise rounding structure (Proposition \ref{prop.reduction}), characterize the  set of optimal solutions (Proposition \ref{prop:largest-fractional-parts}), derive a relative-error tie-breaking rule (Corollary \ref{corr.tiebreak}), extend the characterization to common-loss separable objectives, and prove weak-majorization optimality (Theorem \ref{thm:weak-majorization-oric}). Section~\ref{sec:algo} presents sorting-based and deterministic linear-time implementations and analyzes their computational complexity. Finally, Section~\ref{sec:comparison} compares the proposed method with commonly used deterministic and randomized rounding procedures.

\section{Optimal rounding under integer constraints}\label{sec:optimalrounding}

\subsection{Notations}
Denote by 
$\bZ$ the set of  integers, 
$\bN$ the set of non-negative integers and
 $\bR$ the set of real numbers.
For a vector $x=(x_1,...,x_N)\in \mathbb{R}^N$ and $q\geq 1$, denote
$$ \|x\|_q= \left( \sum_{i=1}^N |x_i|^q \right)^{1/q}$$
  For a real number $y\in \bR$,  we denote 
$$ \lfloor y\rfloor=\sup\{m\in \bZ, \quad m\leq y\} \quad \lceil y\rceil=\inf\{m\in \bZ, \quad m\geq y\} $$
For  $(z_1,z_2,..,z_N)\in \mathbb{R}^N$ we denote the {\it ordered} sequence as
\[
z_{(1)}\geq\cdots\geq z_{(N)}.
\]
\subsection{Problem set-up} 
The problem of optimal rounding   under integer constraints can be formulated in the following way:
\begin{problem}[Optimal rounding under integer constraints ({\bf ORIC})\ ]\label{problem:1} \ \\
Let $q\geq 1$.
Given positive real numbers $x=(x_1,...,x_N)\in \mathbb{R}_+^N$ with 
\ba \sum_{i=1}^N  x_i= M\in \bN  \label{eq:constraint1}\ea
find a set of integers $m=(m_1,...,m_N)\in \mathbb{N}^N$ which minimizes
\ba \|x-m\|_q  =\left(\sum_{i=1}^N |x_i-m_i|^q\right)^{1/q} \quad{\rm under}\qquad\sum_{i=1}^N  m_i= M.  \label{eq:Lq}\ea
\end{problem}
We denote the corresponding  ($L^q$) rounding error 
\ba  V_q(x)=\inf\{ \|x-m\|_q , \qquad m\in \bN^N, \sum_{i=1}^N m_i= M\}. \quad \label{eq:Vq}\ea
The non-trivial feature of the problem is  the presence of the integer constraint.
As the example $x=(2.25, 3.4, 4.35)$ shows, componentwise rounding to the nearest integer may fail to satisfy such a constraint. 

Problem \ref{problem:1} is a special case of the following integer programming problem.
\begin{problem}\label{problem:2}
Given a convex function $f:\bR^N \to \bR $ 
find a set of integers $m=(m_1,...,m_N)\in \mathbb{N}^N$ which minimizes
\ba \inf_{m\in \bN^N} f(m) \quad {\rm under}\qquad\sum_{i=1}^N  m_i= M\in \bN. \label{eq:f}\ea
\end{problem}
This is an optimization problem over the finite set
$$\{ m\in \bN^N, \sum_{i=1}^N m_i= M\}$$
 so the infimum is always attained.
We denote by
\ba  V(f)=\min \{ f(m) , \qquad m\in \bN^N, \sum_{i=1}^N m_i= M\} \quad \label{eq:V}\ea
 the value of this  minimum.
 The feasible set has cardinality
\[
\binom{M+N-1}{N-1},
\]
which can grow exponentially with \(N\) when \(M\) grows proportionally
with \(N\).
So,
at first sight, Problems \ref{problem:1} and  \ref{problem:2}  appear to be  integer optimization problems with exponential complexity. 

Our contribution is to study the structure of 
these problems and show how some of them can be solved using an algorithm with polynomial -in fact {\it linear}- complexity in $N$. 
We propose such a solution for Problem~\ref{problem:1} and the common-loss separable
subclass of Problem~\ref{problem:2}, described in Proposition~\ref{prop:2}.

We first show in Section \ref{sec:main} that, notwithstanding the constraint, the solution necessarily consists in rounding each component either up or down (Proposition \ref{prop.reduction}). We can thus  reformulate the problem as an optimization problem on $\{0,1\}^N$. 
  Next, in Section~\ref{sec:algo}  we first describe a sorting-based implementation
with \(O(N\log N)\) complexity and then give a deterministic
linear-time implementation with \(O(N)\) worst-case complexity.

\subsection{Related problems and ramifications} 
\label{sec.problems}
Problem \ref{problem:1} is a 'pure integer programming' problem in the sense that the relaxation to the case where $m\in \bR^N$ is trivial. As such, it enters as a building block in many integer and mixed-integer programming problems in which one first solves a relaxation  of the problem to real variables then projects back the solution of the relaxed problem  onto $\mathbb{Z}^N$.

The following problem arises e.g. in rounding problems encountered in accounting, where one rounds $N$ entries while leaving the total unchanged up to the nearest dollar:
\begin{problem}[Decimal rounding with exact preservation of total]
\label{prob:decimal-rounding}
Let $q \geq 1$. Given
$
x=(x_1,\ldots,x_N)\in\mathbb{R}_+^N
$
and $k\in \bN$ such that
$$
10^k\sum_{i=1}^N x_i=M\in\mathbb{N},
$$
find
$
y=(y_1,\ldots,y_N)\in 10^{-k}\mathbb{N}^N
$
which minimizes
\[
\sum_{i=1}^N |x_i-y_i|^q
\quad
{\rm subject\  to}\qquad
\sum_{i=1}^N y_i=\sum_{i=1}^N x_i
=\frac{M}{10^k}.
\]
\end{problem}
This problem is equivalent to Problem~1. Indeed, define
$$ X_i=10^k x_i
\qquad\text{and}\qquad
m_i=10^k y_i.$$
Then $m\in\mathbb{N}^N$ and the constraint becomes
$$
\sum_{i=1}^N m_i=M=\sum_{i=1}^N X_i.$$
Moreover,
$$
\sum_{i=1}^N |x_i-y_i|^q=
10^{-kq}\sum_{i=1}^N |X_i-m_i|^q.
$$
So minimizing this objective is equivalent to minimizing
$$
\sum_{i=1}^N |X_i-m_i|^q\quad 
{\rm under}
\quad m\in\mathbb{N}^N,
\qquad
\sum_{i=1}^N m_i=M,$$
which is precisely Problem~\ref{problem:1} applied to $X=10^k x$.

Problem \ref{problem:1} has multiple solutions precisely when there is a tie at the
rounding cutoff, that is, when
\[
1\leq I\leq N-1
\qquad\text{and}\qquad
z_{(I)}=z_{(I+1)},
\]
where \(z_{(1)}\geq\cdots\geq z_{(N)}\) are the fractional parts in
decreasing order.
In this case one might consider minimizing the relative rounding error among all solutions of Problem \ref{problem:1}: 
\begin{problem}[Optimal rounding with smallest relative error]\label{problem:4}
Let $x_i>0$ and
\[
\mathcal{M}_q(x)
=
\operatorname*{arg\,min}_{\substack{m\in\mathbb{N}^N\\
                                    \sum_{i=1}^N m_i=M}}
\sum_{i=1}^N |x_i-m_i|^q.
\]
Find \(m^*\in\mathcal{M}_q(x)\) such that
\[
\sum_{i=1}^N\frac{|x_i-m_i^*|^q}{x_i^q}
=
\min_{m\in\mathcal{M}_q(x)}
\sum_{i=1}^N\frac{|x_i-m_i|^q}{x_i^q}.
\]
\end{problem}


\subsection{Applications and scope}\label{sec:applications}
Problem~\ref{problem:1} arises   whenever \(M\) identical, indivisible units
must be distributed among \(N\) recipients according to a real-valued
target allocation
\[
x=(x_1,\ldots,x_N),
\qquad
x_i\geq0,
\qquad
\sum_{i=1}^N x_i=M.
\]
The objective is to construct an integer allocation \(m\) that preserves
the total number of units while remaining as close as possible to the
target allocation.

\begin{example}[Apportionment \cite{Pukelsheim2014}]
{\rm Let \(v_i>0\) denote the number of votes received by party \(i\), or
the population associated with constituency \(i\), and let \(M\) be
the number of seats to be allocated. The exact quota is
\[
x_i=M\frac{v_i}{\sum_{j=1}^N v_j},
\qquad
\sum_{i=1}^N x_i=M.
\]
Problem~\ref{problem:1} asks for an integer seat allocation \(m\) satisfying
\[
\sum_{i=1}^N m_i=M
\]
and minimizing the discrepancy from the exact quota vector. The ORIC
allocation detailed in Section \ref{sec:main} coincides with the largest-remainder, or Hare--Niemeyer \cite{niemeyer2008}, apportionment. The optimization properties established below concern
closeness to the quota vector; other axiomatic properties of
apportionment methods constitute separate criteria.

The same mathematical problem arises when election seats are to be apportioned across
constituencies according to their populations. See Balinski and Young \cite{balinski} for a detailed exposition
and Grimmett \cite{grimmett2012} for a discussion in the context of the European Parliament.
This problem has a long history in mathematics, dating back to Polya \cite{polya1919,polya1919b}, and continues to generate a lot of interest \cite{niemeyer2008,grimmett2012,janson2014}.}
\label{ex:voting}
\end{example}

\begin{example}[Allocation of identical units]
Suppose that \(M\) identical production, storage, computational, or
investment units are to be allocated among \(N\) activities. If
\(x_i\) denotes the desired real-valued allocation to activity \(i\),
Problem~1 gives an integer allocation preserving the total number of
units and minimizing the rounding error.
\end{example}

The method also applies as a rounding subroutine when the residual
integer feasibility condition of a relaxed optimization problem reduces
to a single conservation or cardinality constraint. General portfolio,
assignment, and scheduling problems typically involve additional
weighted, capacity, or coupling constraints and therefore do not reduce
directly to Problem~\ref{problem:1}.

\section{Structure and optimality of  solutions}\label{sec:main}

We first establish that every solution to Problem~\ref{problem:1} rounds each coordinate either to its floor or to its ceiling. We then characterize the complete set of optimal solutions through the largest fractional parts and derive a relative-error tie-breaking rule when the optimum is not unique. Next, we extend the largest-fractional-parts characterization to the common-loss separable subclass of Problem~\ref{problem:2}, considered in Proposition~\ref{prop:2}. Finally, we show that the absolute-error vector of an ORIC solution is minimal in the weak-majorization order.

\subsection{Reduction to componentwise rounding}
\begin{proposition}[Reduction to term-by-term rounding]\label{prop.reduction}\ \\
Let $x\notin \bN^N$ and $m^*=(m^*_1,...,m^*_N)\in \mathbb{N}^N$ be a solution 
to Problem \ref{problem:1}. Then
\ba \forall i=1..N,\qquad m^*_i\in \{\lfloor x_i\rfloor,\lceil x_i\rceil\}.\ea
\end{proposition}
\begin{proof}
    Define the feasible set $\mathcal F=
\left\{ m\in\mathbb N^N:
\sum_{i=1}^N m_i=M
\right\}$ and
\[
    F_q(m)=\sum_{i=1}^N |m_i-x_i|^q.
\]
As the feasible set $\mathcal{F}$ is non-empty and finite, $F_q$ attains its minimum and the set of minimizers is non-empty.
We now show that every minimizer $m$ rounds each coordinate either downward or
upward. Suppose that a minimizer $m$ has a coordinate satisfying
$
    m_i<\lfloor x_i\rfloor.$
Because
\[
    \sum_{\ell=1}^N (m_\ell-x_\ell)=0,
\]
there exists an index $j$ such that $m_j>x_j$. Set
\[
    A=x_i-m_i\ge 1,
    \qquad
    B=m_j-x_j>0,
\]
and define
$
    m'=m+e_i-e_j,$
where $e_i$ denotes the $i$th standard basis vector. Then $m'\in\mathcal{F}$.
Moreover,
\begin{align*}
    F_q(m)-F_q(m')
    &=A^q-(A-1)^q+B^q-|B-1|^q.
\end{align*}
Since $A\ge1$ and $q\ge1$,
\[
    A^q-(A-1)^q\ge 1.
\]
If $0<B<1$, then
\[
    B^q-|B-1|^q=B^q-(1-B)^q>-1,
\]
whereas if $B\ge1$, then
\[
    B^q-|B-1|^q\ge 0.
\]
Hence $F_q(m)-F_q(m')>0$, contradicting the optimality of $m$.
Therefore no minimizer can satisfy $m_i<\lfloor x_i\rfloor$.
A symmetric argument shows that no minimizer can satisfy
$m_i>\lceil x_i\rceil$.\end{proof}

Proposition~\ref{prop.reduction}  therefore reduces Problem~\ref{problem:1} to a binary optimization
problem subject to the cardinality constraint that exactly
\[
I=M-\sum_{i=1}^N\lfloor x_i\rfloor
\]
noninteger components are rounded upward.
\subsection{Characterization by largest fractional parts}
\label{subsec:largest-fractional-parts}

Define
$
z_i=x_i-\lfloor x_i\rfloor\in[0,1),
$
and 
\[
I=M-\sum_{i=1}^N \lfloor x_i\rfloor
  =\sum_{i=1}^N z_i\in\mathbb{N}.
\]
Since \(z_i<1\) for every \(i\), we have \(0\leq I\leq N-1\).
By Proposition~3.1, every solution to Problem~1 rounds each component
either downward or upward. Hence every solution has the form
\[
m_i= \lfloor x_i\rfloor +1_{\{i\in S\}},
\qquad i=1,\ldots,N,
\]
for some \(S\subseteq\{i:z_i>0\}\). The sum constraint then requires
$ |S|=I.$
Thus, the constrained rounding problem reduces to selecting the \(I\)
coordinates to be rounded upward.
The following result characterizes all solutions of Problem \ref{problem:1}.
\begin{proposition}[Characterization by largest fractional parts]
\label{prop:largest-fractional-parts}\ \\
Let \(q\geq 1\), and let
\[
x=(x_1,\ldots,x_N)\in\mathbb{R}_+^N,
\qquad
\sum_{i=1}^N x_i=M\in\mathbb{N}.
\]
For each \(i\), write \(a_i=\lfloor x_i\rfloor\) and
\(z_i=x_i-a_i\), and let
\[
I=M-\sum_{i=1}^N a_i=\sum_{i=1}^N z_i.
\]
Let \(S^*\subseteq\{1,\ldots,N\}\) consist of \(I\) indices having the
largest fractional parts \(z_i\), with ties at the cutoff resolved
arbitrarily, and define
\[
m_i^*=a_i+1_{\{i\in S^*\}},
\qquad i=1,\ldots,N.
\]
Then:
\begin{enumerate}
\item \(m^*\in\mathbb{N}^N\) and
\[
\sum_{i=1}^N m_i^*=M.
\]
\item The vector \(m^*\) solves Problem~1:
\[
\sum_{i=1}^N |x_i-m_i^*|^q
=
\min_{\substack{m\in\mathbb{N}^N\\
                 \sum_{i=1}^N m_i=M}}
\sum_{i=1}^N |x_i-m_i|^q.
\]
The same choice of \(S^*\) is optimal for every \(q\geq1\).
\item The set of solutions to Problem~1 consists precisely of
the vectors obtained by rounding upward \(I\) coordinates whose
fractional parts are among the \(I\) largest fractional parts.
\item The solution is unique if and only if either \(I=0\), or $ z_{(I)}>z_{(I+1)},$
where \(z_{(1)}\geq\cdots\geq z_{(N)}\) are the fractional parts in
decreasing order.
\end{enumerate}
\end{proposition}

\begin{proof}
The feasibility statement follows directly from the definition of \(m^*\):
\[
\sum_{i=1}^N m_i^*
=
\sum_{i=1}^N a_i+|S^*|
=
\sum_{i=1}^N a_i+I
=
M.
\]

By Proposition~3.1, it is sufficient to consider vectors of the form
\[
m_i=a_i+1_{\{i\in S\}},
\qquad |S|=I.
\]
For any such set \(S\),
\begin{align*}
\sum_{i=1}^N |x_i-m_i|^q
&=
\sum_{i\notin S} z_i^q
+
\sum_{i\in S}(1-z_i)^q\\
&=
\sum_{i=1}^N z_i^q
+
\sum_{i\in S}\bigl((1-z_i)^q-z_i^q\bigr).
\end{align*}
Define the incremental cost of rounding a coordinate upward by
\[
d_q(z)=(1-z)^q-z^q,
\qquad 0\leq z\leq1.
\]
For \(q=1\),
\[
d_1(z)=1-2z,
\]
which is strictly decreasing. For \(q>1\),
\[
d_q'(z)
=
-q(1-z)^{q-1}-qz^{q-1}<0,
\qquad 0<z<1.
\]
Thus \(d_q\) is strictly decreasing on \([0,1]\) for every \(q\geq1\).
Minimizing the objective over all sets \(S\) of cardinality \(I\) is
therefore equivalent to selecting the \(I\) smallest values of
\(d_q(z_i)\), which are precisely the \(I\) largest values of \(z_i\).
This proves the optimality of \(m^*\). Since the ordering induced by
\(d_q\) is the ordering of the fractional parts for every \(q\geq1\),
the same upward-rounded set is simultaneously optimal for all such \(q\).

The same argument also characterizes all solutions. Every coordinate with
fractional part strictly above the cutoff must be rounded upward, every
coordinate with fractional part strictly below the cutoff must be rounded
downward, and the required number of coordinates at the cutoff may be
chosen arbitrarily. Hence the solution is unique when \(I=0\), or when
there is a strict separation \(z_{(I)}>z_{(I+1)}\). If
\(z_{(I)}=z_{(I+1)}\), different choices within the cutoff tie set yield
distinct solutions with the same objective value.
\end{proof}
We will call 'ORIC solution' to refer to the set of solutions characterized by the result (3) above:
\begin{definition}[ORIC solution]
We call     \emph{ORIC solution} of Problem \ref{problem:1} any vector $m\in \bN^N$ obtained by rounding upward \(I\) coordinates of $x$ having
the \(I\) largest fractional parts, where $I=M-\sum_{i=1}^N \lfloor x_i\rfloor $. 
\end{definition}
This definition
is independent of \(q\geq1\).

\subsection{Relative error tie-breaking}
\label{sec.relativerror}
\begin{corollary}[Relative-error tie-breaking]
\label{corr.tiebreak} 
Under the assumptions of Proposition \ref{prop:largest-fractional-parts}, suppose that  one wishes to minimize
\[
R_q(m)=\sum_{i=1}^N\frac{|x_i-m_i|^q}{x_i^q}.
\]
among the solutions of Problem~\ref{problem:1}.
Assume \(I\geq1\), and let \(t=z_{(I)}\) be the fractional
part at the cutoff. Define
\[
    T=\{i:z_i=t\},
    \qquad
    k=I-\#\{i:z_i>t\}.
\]
Thus exactly $k$ indices in $T$ must be rounded upward. A minimizer of \(R_q\) among the solutions of Problem~\ref{problem:1} is obtained
by choosing
\[
\begin{cases}
\text{the $k$ largest values of $x_i$ in $T$}, & t<\tfrac12,\\[1mm]
\text{any $k$ indices in $T$}, & t=\tfrac12,\\[1mm]
\text{the $k$ smallest values of $x_i$ in $T$}, & t>\tfrac12.
\end{cases}
\]
Since all indices in $T$ have the same fractional part, ordering by $x_i$ is
equivalent to ordering by $\lfloor x_i\rfloor$.
\end{corollary}

\begin{proof} Among the solutions of Problem~\ref{problem:1}, only the choice of indices inside
the cutoff tie set \(T\) is variable.
For every $i\in T$, the change in relative error caused by
rounding $i$ upward rather than downward is
\[
    \Delta_i
    =
    \frac{(1-t)^q-t^q}{x_i^q}.
\]
Exactly $k$ such increments must be selected.

If $t<\tfrac12$, then $(1-t)^q-t^q>0$, so the $k$ smallest increments are
obtained by choosing the $k$ largest values of $x_i$. If $t>\tfrac12$, then
$(1-t)^q-t^q<0$, so the smallest, that is, most negative, increments are
obtained by choosing the $k$ smallest values of $x_i$. If $t=\tfrac12$, every
increment is zero, so every choice of $k$ tied coordinates has the same relative
error. 
\end{proof}

\subsection{Extension to common-loss separable objectives}
We now turn to Problem \ref{problem:2}.  The following result gives a solution to Problem \ref{problem:2} for 'common-loss, separable objectives' i.e. when $f$ has the form 
$$f(y)= C+ \sum_{i=1}^N \phi(y_i-x_i)$$
on the feasible hyperplane $\sum_{i=1}^N y_i=M$.
\begin{proposition}\label{prop:2}
Let $x^*=(x^*_1,...,x^*_N)\in \mathbb{R}_+^N$ such that  $\sum_{i=1}^N x^*_i=M\in \mathbb{N}$.
Let $\phi:\mathbb{R}\mapsto \mathbb{R}$ be strictly convex and define
$$f(y)=\sum_{i=1}^N \phi(y_i-x_i^*).$$
Then $x^*$ is the unique solution of
\ba \inf_{y\in \bR_+^N} f(y) \quad {\rm under} \qquad \sum_{i=1}^N y_i=M  \label{eq:fmin}\ea
and every solution $m^*=(m^*_1,...,m^*_N)\in\mathbb{N}^N$ of 
    \ba\inf_{m\in \bN^N} f(m) \quad {\rm under} \qquad \sum_{i=1}^N m_i=M.   \label{eq.2int}\ea
satisfies  
\ba \forall i\in [1,N],\qquad m^*_i\in \{\lfloor x^*_i\rfloor ,\lceil x_i^*\rceil \}.\ea 
Let $z_i=x^*_i-\lfloor x^*_i\rfloor $ and
$$ I=\sum_{i=1}^N z_i= M- \sum_{i=1}^N \lfloor x^*_i\rfloor \in \mathbb{N}.$$
Then an integer minimizer is obtained by rounding upward exactly the $I$ components with the largest fractional parts $z_i$, with cutoffs resolved arbitrarily.
\end{proposition}
\begin{proof}
For any feasible $y$ let $u_i=y_i-x_i^*$. we have 
$\sum_i u_i=0.$ Jensen's inequality gives
$$\frac{1}{N}\sum_{i=1}^N \phi(u_i)\geq \phi( \frac{1}{N}\sum_{i=1}^N u_i)=\phi(0).$$
Strict convexity gives equality only when $u_1=...=u_N$. Since $\sum_{i=1}^N u_i=0$ this implies $u=0$. Hence $x^*$ is the unique minimizer for the constrained optimization problem \eqref{eq:fmin}.

Let $m$ be a minimizer in \eqref{eq.2int} and $d_i=m_i-x_i^*.$ Suppose $m_i<\lfloor x_i^*\rfloor.$ Then $d_i\leq -1$. Since $\sum d_i=0$ there exists $j\neq i$ with $d_j>0$.
The unit increment function $h(t)=\phi(t+1)-\phi(t)$ is strictly increasing. Because $d_j-d_i>1,$ $h(d_i)< h(d_j-1).$
Consequently
$$f(m+e_i-e_j)-f(m)=\phi(d_i+1)-\phi(d_i)+ \phi(d_j-1)-\phi(d_j)=h(d_i)-h(d_{j}-1)<0$$
which contradicts optimality.
The same exchange argument   with the direction of the transfer
reversed, excludes $m_i>\lceil x_i^*\rceil.$ Therefore every coordinate is rounded either up or down.
 
Rounding coordinate $i$ upward changes the objective by 
$$\delta_i= \phi(1-z_i)-\phi(-z_i)=h(-z_i)$$
Since $h$ is strictly increasing, $\delta_i$ is strictly decreasing in $z_i$. Thus the $I$ smallest incremental costs correspond exactly to the $I$ largest fractional parts.
\end{proof}

Taking $\phi(u)=|u|^q$ for $q>1$ corresponds to the $L^q$ minimization problem under integer constraints:
\ba \inf_{m\in \bN^N} \sum_{i=1}^N |m_i-x_i^*|^q \quad {\rm under} \qquad \sum_{i=1}^N m_i=M\in \mathbb{N}  \label{eq:Lqint}\ea
\subsection{Weak majorization and universal optimality}
\label{subsec:majorization}

We now show that the optimality property of the ORIC solution is
stronger than simultaneous \(L^q\)-optimality. In fact, its vector of
absolute rounding errors is minimal in the weak-majorization order \cite{MarshallOlkinArnold}.
For \(v=(v_1,\ldots,v_N)\in\mathbb{R}_+^N\),  define \(T_0(v)=0\) and
\[
T_k(v)=\sum_{r=1}^k v_{(r)},
\qquad k=1,\ldots,N,\quad{\rm where}\quad v_{(1)}\geq v_{(2)}\geq\cdots\geq v_{(N)}.
\]
\begin{definition}[Weak submajorization]
\label{def:weak-submajorization}
For \(u,v\in\mathbb{R}_+^N\), we say that \(u\) is weakly
submajorized by $v$, and write
$
u\preceq_{\mathrm w}v,$
if
\[
T_k(u)\leq T_k(v),
\qquad k=1,\ldots,N.
\]
Thus \(u\preceq_{\mathrm w}v\) means: for every \(k\), the sum
of the \(k\) largest components of \(u\) is no larger than the sum
of the \(k\) largest components of \(v\).
\end{definition}

For a feasible integer vector \(m\), define its absolute-error vector by
\[
e(m)=
\bigl(
|x_1-m_1|,\ldots,|x_N-m_N|
\bigr).
\]

We first record two elementary observations.

\begin{lemma}[Embedding property]
\label{lem:majorization-embedding}
Suppose that \(u,v\in\mathbb{R}_+^r\) satisfy
\[
u\preceq_{\mathrm w}v.
\]
Then, for every \(w\in\mathbb{R}_+^s\),
$
(u,w)\preceq_{\mathrm w}(v,w),
$
where \((u,w)\) denotes the concatenation of \(u\) and \(w\).
\end{lemma}

\begin{proof}
For every \(k\in\{0,\ldots,r+s\}\), the sum of the \(k\) largest
components of \((u,w)\) may be written as
\[
T_k(u,w)
=
\max_{\ell}
\left\{
T_\ell(u)+T_{k-\ell}(w)
\right\},
\]
where the maximum is over the integers \(\ell\) satisfying
$
0\leq \ell\leq r,
\quad
0\leq k-\ell\leq s.$
Since \(T_\ell(u)\leq T_\ell(v)\) for every \(\ell\), it follows that
\[
\begin{aligned}
T_k(u,w)
&=
\max_{\ell}
\left\{
T_\ell(u)+T_{k-\ell}(w)
\right\} \\
&\leq
\max_{\ell}
\left\{
T_\ell(v)+T_{k-\ell}(w)
\right\}
=
T_k(v,w).
\end{aligned}
\]
Hence
$
(u,w)\preceq_{\mathrm w}(v,w).$\end{proof}

\begin{lemma}[Two-coordinate improvements]
\label{lem:two-coordinate-majorization}
The following weak-majorization relations hold.

\begin{enumerate}
\item If \(A\geq1\) and \(B>0\), then
      \[
      \bigl(A-1,|B-1|\bigr)
      \preceq_{\mathrm w}
      (A,B),\qquad 
      A-1+|B-1|<A+B.
      \]
\item If \(0\leq b<a<1\), then
      \[
      (1-a,b)
      \preceq_{\mathrm w}
      (a,1-b),
      \qquad
      1-a+b<a+1-b.
      \]
\end{enumerate}
\end{lemma}

\begin{proof}
For two-dimensional nonnegative vectors, weak submajorization is
equivalent to comparing both the maximum component and the sum of
the two components.
For the first assertion, if \(B\geq1\), then
\[
A-1\leq A,
\qquad
|B-1|=B-1\leq B.
\]
If \(0<B<1\), then
\[
|B-1|=1-B<1\leq A.
\]
In either case,
\[
\max\{A-1,|B-1|\}
\leq
\max\{A,B\}.
\]
Moreover, if \(0<B<1\), then
\[
A-1+|B-1|
=
A-B
<
A+B,
\]
whereas if \(B\geq1\), then
\[
A-1+|B-1|
=
A+B-2
<
A+B.
\]
This proves the first assertion.

For the second assertion, \(b<a\) implies
\[
b<a
\qquad\text{and}\qquad
1-a<1-b.
\]
Consequently,
\[
\max\{1-a,b\}
\leq
\max\{a,1-b\}.
\]
Also,
\[
1-a+b
=
1-(a-b)
<
1+(a-b)
=
a+1-b.
\]
This proves the second assertion.
\end{proof}

We may now state the principal majorization result.

\begin{theorem}[Weak-majorization optimality of ORIC]
\label{thm:weak-majorization-oric}
Let
\[
x=(x_1,\ldots,x_N)\in\mathbb{R}_+^N,
\qquad
\sum_{i=1}^N x_i=M\in\mathbb{N},
\]
and let
\[
\mathcal F=
\left\{
m\in\mathbb{N}^N:
\sum_{i=1}^N m_i=M
\right\}.
\]
For each \(i\), write
$
a_i=\lfloor x_i\rfloor,
\qquad
z_i=x_i-a_i,
$
and define
\[
I=M-\sum_{i=1}^N a_i
  =\sum_{i=1}^N z_i.
\]

Let \(m^*\) be any ORIC solution:
$
m_i^*
=
a_i+\mathbf{1}_{\{i\in S^*\}},$
where \(S^*\) consists of \(I\) indices having the largest fractional
parts \(z_i\), with ties at the cutoff resolved arbitrarily.
Then, for every \(m\in\mathcal F\),
\[
e(m^*)\preceq_{\mathrm w}e(m).
\]
Equivalently,
\[
\sum_{r=1}^k e_{(r)}(m^*)
\leq
\sum_{r=1}^k e_{(r)}(m),
\qquad k=1,\ldots,N.
\]

Moreover,
\[
\sum_{i=1}^N |x_i-m_i^*|
<
\sum_{i=1}^N |x_i-m_i|
\]
for every feasible \(m\) which is not an ORIC solution.
Consequently, equality in the weak-majorization inequality for
\(k=N\) holds if and only if \(m\) is an ORIC solution.
\end{theorem}

\begin{proof}
Fix \(m\in\mathcal F\). We transform \(m\) into an ORIC solution
through a finite sequence of feasible exchanges. At every exchange,
the new error vector is weakly submajorized by the preceding error
vector.

If \(I=0\), then
\[
\sum_{i=1}^N z_i=0.
\]
Since \(z_i\geq0\), it follows that \(z_i=0\) for every \(i\), so
\(x\in\mathbb N^N\) and \(m^*=x\). Hence
\[
e(m^*)=0\preceq_{\mathrm w}e(m)
\]
for every \(m\in\mathcal F\). Moreover, if \(m\neq m^*\), then
\[
\sum_{i=1}^N|x_i-m_i|>0
=
\sum_{i=1}^N|x_i-m_i^*|.
\]
We may therefore assume below that \(I\geq1\). 
Suppose first that
$
m_i<\lfloor x_i\rfloor
$ for some \(i\). Since
\[
\sum_{\ell=1}^N(m_\ell-x_\ell)=0,
\]
there exists an index \(j\) such that
$
m_j>x_j.$
Let
\[
A=x_i-m_i\geq1,
\qquad
B=m_j-x_j>0,
\]
and define
$
m'=m+e_i-e_j.$
Since \(m_j>x_j\geq0\) and \(m_j\in\mathbb N\), we have
\(m_j\geq1\). Hence \(m_j-1\in\mathbb N\), and therefore
\(m'\in\mathbb N^N\). We also have
\[
\sum_{\ell=1}^N m_\ell'
=
\sum_{\ell=1}^N m_\ell
=
M.
\]
Thus \(m'\in\mathcal F\).
The two affected errors change from
$
(A,B)$
to $
(A-1,|B-1|).$
Lemma~\ref{lem:two-coordinate-majorization} gives
\[
(A-1,|B-1|)
\preceq_{\mathrm w}
(A,B).
\]
The remaining \(N-2\) errors are unchanged. By
Lemma~\ref{lem:majorization-embedding},
$
e(m')\preceq_{\mathrm w}e(m).$
In addition,
\[
\sum_{\ell=1}^N |x_\ell-m_\ell'|
<
\sum_{\ell=1}^N |x_\ell-m_\ell|.
\]

Suppose instead that
$
m_j>\lceil x_j\rceil $
for some \(j\). Since the deviations still sum to zero, there exists
an index \(i\) such that
$
m_i<x_i.$
Set
\[
A=m_j-x_j\geq1,
\qquad
B=x_i-m_i>0,
\]
and again define
\[
m'=m+e_i-e_j.
\]
The two affected errors change from
$(A,B)$
to
$
(A-1,|B-1|).$
The same argument therefore gives
\[
e(m')\preceq_{\mathrm w}e(m)
\]
and a strict reduction of the total absolute error.

We repeat these exchanges whenever a coordinate lies below its floor
or above its ceiling. Each exchange preserves feasibility and strictly
decreases
\[
\sum_{i=1}^N |x_i-m_i|.
\]
Since the feasible set \(\mathcal F\) is finite, the procedure must
terminate. At termination, the resulting feasible vector
\(\widetilde m\) satisfies
\[
\widetilde m_i
\in
\{\lfloor x_i\rfloor,\lceil x_i\rceil\},
\qquad i=1,\ldots,N,
\]
and
$
e(\widetilde m)\preceq_{\mathrm w}e(m).
$
Therefore
$
\widetilde m_i
=
a_i+\mathbf{1}_{\{i\in S\}}
$
for some set
$
S\subseteq\{i:z_i>0\}.$
The sum constraint gives
\[
M
=
\sum_{i=1}^N a_i+|S|,
\]
so $|S|=I.$
Now suppose that \(S\) does not consist of \(I\) indices having the largest
fractional parts. Then there exist indices
\[
i\notin S,
\qquad
j\in S,
\]
such that
$
z_i>z_j.$
The coordinate \(i\) is currently rounded downward and has error
\(z_i\), while coordinate \(j\) is currently rounded upward and has
error \(1-z_j\).

Define a new upward-rounded set by
\[
S'=(S\setminus\{j\})\cup\{i\}.
\]
The corresponding feasible vector \(\widetilde m'\) is obtained by
rounding \(i\) upward and \(j\) downward. The affected pair of errors
changes from
$
(z_i,1-z_j)
$
to
$
(1-z_i,z_j).
$
Since
\[
0\leq z_j<z_i<1,
\]
Lemma~\ref{lem:two-coordinate-majorization} gives
\[
(1-z_i,z_j)
\preceq_{\mathrm w}
(z_i,1-z_j).
\]
After adjoining the unchanged errors, we obtain
$ e(\widetilde m')
\preceq_{\mathrm w}
e(\widetilde m).$
Furthermore,
\[
\begin{aligned}
(1-z_i)+z_j
&=
1-(z_i-z_j) \\
&<
1+(z_i-z_j) =
z_i+(1-z_j),
\end{aligned}
\]
so the total absolute error decreases strictly.

We repeat this exchange whenever such a pair \(i\notin S\), \(j\in S\)
with \(z_i>z_j\) exists. Since there are finitely many subsets of
cardinality \(I\), the procedure terminates. At termination, every
selected fractional part is at least as large as every unselected
fractional part. Hence the final selected set consists of \(I\)
indices having the largest fractional parts. The resulting vector
\(\widehat m\) is therefore an ORIC solution.
By transitivity of weak submajorization,
\[
e(\widehat m)
\preceq_{\mathrm w}
e(m).
\]
It remains only to observe that all ORIC solutions have the same
multiset of absolute errors. Indeed, let
$
t=z_{(I)} $
be the cutoff fractional part, and define
\[
H=\{i:z_i>t\},
\qquad
T=\{i:z_i=t\},
\qquad
L=\{i:z_i<t\}.
\]
Every ORIC solution rounds every index in \(H\) upward, every index in
\(L\) downward, and exactly
$I-|H| $
indices in \(T\) upward. The errors contributed by the cutoff tie set
are therefore always
$
I-|H|
\quad\text{copies of }1-t$
and
$|T|-(I-|H|)
\quad\text{copies of }t.$
Thus every ORIC solution has the same decreasingly ordered error
vector. In particular,
\[
e_{(r)}(m^*)=e_{(r)}(\widehat m),
\qquad r=1,\ldots,N.
\]
It follows that
$
e(m^*)\preceq_{\mathrm w}e(m).$
Finally, if \(m\) is not an ORIC solution, then at least one exchange
in the steps above is required. Every such exchange strictly decreases
the total absolute error. Hence
\[
\sum_{i=1}^N |x_i-m_i^*|
<
\sum_{i=1}^N |x_i-m_i|.
\]
This completes the proof. \end{proof}
The weak-majorization theorem yields a collection of simultaneous
optimality results.
\begin{corollary}[Universal optimality]
\label{cor:universal-optimality}
Under the assumptions of
Theorem~\ref{thm:weak-majorization-oric}, let \(m^*\) be any ORIC
solution. Then the following assertions hold.
\begin{enumerate}
\item For every \(k=1,\ldots,N\), \(m^*\) minimizes the sum of the
      \(k\) largest absolute rounding errors:
      \[
      \sum_{r=1}^k e_{(r)}(m^*)
      =
      \min_{m\in\mathcal F}
      \sum_{r=1}^k e_{(r)}(m).
      \]

\item For every convex nondecreasing function
      $
      \psi:[0,\infty)\longrightarrow\mathbb{R},
      $
      \(m^*\) minimizes the separable loss
      $
      \sum_{i=1}^N
      \psi\bigl(|x_i-m_i|\bigr):
      $
      \[
\forall m\in\mathcal F,\qquad      \sum_{i=1}^N
      \psi\bigl(|x_i-m_i^*|\bigr)
      \leq
      \sum_{i=1}^N
      \psi\bigl(|x_i-m_i|\bigr).
      \]
\item  \(m^*\) minimizes every symmetric, convex,
      coordinatewise nondecreasing function of the absolute-error
      vector.
\item In particular, \(m^*\) simultaneously minimizes
      \[
      \|x-m\|_q,
      \qquad 1\leq q\leq\infty,
      \]
      and every symmetric gauge norm of the error vector.
\end{enumerate}
\end{corollary}
\begin{proof}
Part 1 is exactly the defining inequality of weak submajorization.
We give a direct proof of Part 2. For \(v\in\mathbb{R}_+^N\) and
\(s\geq0\), define
\[
H_s(v)
=
\sum_{i=1}^N(v_i-s)_+,
\qquad
(r)_+=\max\{r,0\}.
\]
If the components of \(v\) are arranged in decreasing order, then
\[
H_s(v)
=
\max_{0\leq k\leq N}
\left\{
T_k(v)-ks
\right\}.
\]
Indeed, the maximum is attained when \(k\) is the number of components
of \(v\) that exceed \(s\).
Since
\[
e(m^*)\preceq_{\mathrm w}e(m),
\]
we have
\[
T_k(e(m^*))\leq T_k(e(m))
\]
for every \(k\). Therefore
\[
\begin{aligned}
H_s(e(m^*))
&=
\max_{0\leq k\leq N}
\left\{
T_k(e(m^*))-ks
\right\} \leq
\max_{0\leq k\leq N}
\left\{
T_k(e(m))-ks
\right\} =
H_s(e(m)).
\end{aligned}
\]
Now suppose first that \(\psi\) is a convex, nondecreasing,
piecewise-linear function. It can be written in the form
\[
\psi(t)
=
c+\alpha t+
\sum_{\ell=1}^J
\beta_\ell(t-s_\ell)_+,
\]
where
$
\alpha\geq0,
\beta_\ell\geq 0.$
Consequently,
\[
\begin{aligned}
\sum_{i=1}^N\psi(e_i(m^*))
&=
Nc
+\alpha T_N(e(m^*))
+\sum_{\ell=1}^J
\beta_\ell H_{s_\ell}(e(m^*)) \\
&\leq
Nc
+\alpha T_N(e(m))
+\sum_{\ell=1}^J
\beta_\ell H_{s_\ell}(e(m)) =
\sum_{i=1}^N\psi(e_i(m)).
\end{aligned}
\]
For a general convex nondecreasing function \(\psi\), choose an interval
\([0,R]\) containing all components of \(e(m^*)\) and \(e(m)\).
Convex nondecreasing piecewise-linear functions obtained by interpolation
on increasingly fine partitions converge uniformly to \(\psi\) on
\([0,R]\). Applying the preceding inequality to these approximations
and passing to the limit proves Part 2.

Part 3: majorization property. If
$
u\preceq_{\mathrm w}v,$
then
$
\Phi(u)\leq\Phi(v)
$
for every symmetric, convex, coordinatewise nondecreasing function
\(\Phi\). Applying this result with
\[
u=e(m^*),
\qquad
v=e(m)
\]
gives the assertion.
For \(1\leq q<\infty\), take
$
\psi(t)=t^q $
in Part 2. Since the function \(r\mapsto r^{1/q}\) is increasing, this
gives
\[
\|x-m^*\|_q\leq\|x-m\|_q.
\]
For \(q=\infty\),
\[
\|x-m^*\|_\infty
=
e_{(1)}(m^*)
\leq
e_{(1)}(m)
=
\|x-m\|_\infty
\]
by Part 1 with \(k=1\).
Finally, the monotonicity of symmetric gauge norms under weak
submajorization is the Ky Fan dominance principle \cite{MarshallOlkinArnold}, which gives the
last assertion.
\end{proof}

\begin{remark}
{\em Theorem~\ref{thm:weak-majorization-oric} strengthens simultaneous \(L^q\)-optimality by showing
that ORIC minimizes every Ky Fan sum of the absolute-error vector and,
more generally, every symmetric convex coordinatewise nondecreasing
loss.
For example, it shows that ORIC
minimizes the largest individual error, the sum of the two largest
errors, and, more generally, the sum of the \(k\) largest errors for
every \(k\), all with the {\it same} integer rounding.}
\end{remark}

\section{Algorithms for optimal rounding under integer constraints}  \label{sec:algo}

Proposition \ref{prop.reduction} reduces the complexity of Problem \ref{problem:1} by confining the search to   the finite set $$\prod_{i=1}^N \{\lfloor x_i\rfloor,\lceil x_i\rceil\}$$ which can be reparameterized as $\{0,1\}^N$ i.e. a componentwise rounding problem. But the size of this set grows exponentially with $N$, so optimization through exhaustive search is not a feasible option, even if $N$ is only moderately large.
\subsection{Implementation via sorting}

We first exhibit a sorting-based algorithm which exploits the structure of the problem, in particular the sum constraint, to compute an optimal solution in $O(N\log N)$ time.
 Given the structure of the objective function  in Problem \ref{problem:1}, the idea is  to optimize  term by term, controlling for the constraint at each step.

We start by rounding all components downwards and compute the constraint shortfall $I= \sum_{i=1}^N  (x_i-m_i)\leq N$. If $I=0$ this means $x$ is already integer valued so one does not need to proceed further. If $I\geq 1$ we need to round upwards exactly $I$ components to meet the constraint. To choose these components  optimally, we 
\begin{enumerate}
\item sort the indices according to decreasing values of  fractional part $x_i-\lfloor x_i\rfloor=x_i-m_i$.
\item For a block of equal fractional parts \(t\), order the
indices by decreasing integer part when \(t<1/2\), by increasing
integer part when \(t>1/2\), and arbitrarily when \(t=1/2\).
\item
In the last step, we proceed to round upwards   the first $I$ components sorted in this order.
\end{enumerate}
Steps 1) and 2) may be done using a QuickSort algorithm \cite{hoare62}.

This yields an algorithm for solving the optimal rounding problem:

{\bf Optimal rounding under integer constraints ({\sc ORIC}):}
\begin{algorithmic}[1]
\STATE Set $\forall i=1..N, m_i=\lfloor x_i\rfloor$.
\STATE Compute constraint shortfall $I(x)= \sum_{i=1}^N  (x_i-m_i)\geq 0.$
\STATE  If $I=0$ then  {\bf end}.
\STATE  Sort indices in decreasing order of fractional part $(x_i-m_i)$:  $$(x_1-m_1)\geq ...\geq (x_N-m_N)\geq 0.$$
\STATE {\bf Tie break}: Sort each block of indices with equal fractional parts $0<t<1$ 
\begin{itemize}
    \item   by decreasing integer part if $t<1/2$,
     \item by increasing integer part if $t>1/2$, and \item arbitrarily
      if $t=1/2$.
\end{itemize}
\FOR{$k=1,\dots,I$}
\STATE   $m_k=\lceil x_k\rceil.$
\ENDFOR
\STATE
\textbf{end} 
\end{algorithmic}
There exists a symmetric version of the algorithm where one initializes with $m_i=\lceil x_i\rceil$ and then needs to round downwards $K-I$ of the components where
$$ K=|\{ i=1..N, z_i>0 \}|$$ It is readily  verified that the two methods yield the same  solutions.

A similar algorithm, based on the sorting of fractional components, but without the sorting step (5), has been used for a long time in the context of   proportional seat allocation 
 for representative assemblies with party list voting systems \cite{balinski}, where it is known as the Hare-Niemeyer or 'largest remainder method'. The focus of the Hare-Niemeyer and related methods is to achieve a 'fair' allocation rather than to minimize a given objective function, so the notion of optimality considered here may or may not be relevant for such applications and more complex considerations apply; we refer to Balinski \& Young \cite{balinski} for a detailed discussion of seat allocation methods in elections. 
Nevertheless, the largest remainder method is a special  case of the method considered here, when all fractional parts are distinct and the analysis of optimality bears many similarities \cite{niemeyer2008}.   

To quantify the complexity of the algorithm, we first note that $$I= \sum_{i=1}^N  (x_i-m_i)\leq N,$$ thus the sorting steps 
~4 and~5 may be implemented using randomized QuickSort \cite{hoare62}  in
expected \(O(N\log N)\) time \cite{hoare62,fredman2014}. Alternatively, a worst-case
\(O(N\log N)\) comparison sort gives a deterministic
\(O(N\log N)\) implementation.

Once the sorting has been done, the rounding of the sorted sequence (Steps 6-8) requires $I\leq N-1$ operations. So, overall, the complexity is dominated by the sorting step: 
\begin{proposition}[ORIC algorithm: implementation by sorting]\ \\
The sorting-based ORIC algorithm returns an ORIC solution and therefore solves Problem \ref{problem:1} for every $q\ge 1$. It has $O(N\log N)$ time complexity and uses $O(N)$ additional space.
\end{proposition}

\begin{proof}
By Proposition~\ref{prop:largest-fractional-parts}, rounding upward the $I$ coordinates with the largest fractional parts produces an ORIC solution and therefore solves Problem~\ref{problem:1} for every $q\geq 1$. Computing $a_i, z_i$, and $I$ requires $O(N)$ operations. Sorting the $N$ indices by decreasing $z_i$ requires $O(N\log N)$ time, and rounding upward the first $I$ coordinates requires at most $N$ additional operations. Hence the overall running time is $O(N\log N)$, with $O(N)$ additional space.

\end{proof}

\subsection{Deterministic implementation with linear complexity}
\label{subsec:linear-time-oric}

The implementation described above sorts all \(N\) fractional parts.
Full sorting is not necessary: it is sufficient to identify the fractional
part at the rounding cutoff and partition the coordinates relative to this
cutoff. We now use this observation to  further reduce the complexity from $O(N\log N)$ to $O(N)$.

We use the following standard selection operation. Given \(n\) real numbers
and an integer \(r\in\{1,\ldots,n\}\), let
\[
\operatorname{Select}_r(v_1,\ldots,v_n)
\]
denote the \(r\)-th largest value among \(v_1,\ldots,v_n\).
The deterministic median-of-medians selection algorithm computes this
order statistic in \(O(n)\) worst-case time; see \cite{BFPRT73}.

For each \(i=1,\ldots,N\), write
$
a_i=\lfloor x_i\rfloor,
\qquad
z_i=x_i-a_i,$
and recall that
\[
I=M-\sum_{i=1}^N a_i
  =\sum_{i=1}^N z_i
  \in\mathbb N.
\]
If \(I\geq 1\), define the cutoff fractional part by
\[
t=\operatorname{Select}_I(z_1,\ldots,z_N)=z_{(I)}.
\]
where $
z_{(1)}\geq\cdots\geq z_{(N)}$
are the fractional parts in decreasing order. Partition the indices into
\[
H=\{i:z_i>t\},
\qquad
T=\{i:z_i=t\},
\qquad
L=\{i:z_i<t\}.
\]
Every index in \(H\) must be rounded upward and every index in \(L\)
must be rounded downward. The number of indices that must be selected
from the cutoff tie set \(T\) is
$
k=I-|H|.$
By the definition of \(t=z_{(I)}\), $
1\leq k\leq |T|.$

This leads to the following deterministic linear-time implementation:
\newpage
\medskip
\noindent
\textbf{Deterministic linear-time optimal rounding under integer
constraints (Linear-ORIC):}

\begin{enumerate}
\item Set
      $
      m_i=\lfloor x_i\rfloor,
      \qquad
      z_i=x_i-m_i,
      \qquad i=1,\ldots,N.
      $
\item Compute the constraint shortfall
      $
      I=\sum_{i=1}^N z_i.
      $
\item If \(I=0\), terminate and return \(m=x\).

\item Compute the cutoff fractional part
      $
      t=\operatorname{Select}_I(z_1,\ldots,z_N).
      $
\item Form the index sets
      \[
      H=\{i:z_i>t\},
      \qquad
      T=\{i:z_i=t\}.
      \]
\item For every \(i\in H\), set
      $
      m_i=\lceil x_i\rceil.$
\item Compute
      $
      k=I-|H|.
      $
\item Select a subset \(C\subseteq T\) with \(|C|=k\) as follows:
      \begin{itemize}
      \item if \(t<1/2\), let \(C\) consist of the \(k\) indices in
            \(T\) having the largest integer parts \(a_i\);
      \item if \(t>1/2\), let \(C\) consist of the \(k\) indices in
            \(T\) having the smallest integer parts \(a_i\);
      \item if \(t=1/2\), choose any \(k\) indices from \(T\).
      \end{itemize}
      Ties in integer parts at the secondary cutoff may be resolved
      arbitrarily.
\item For every \(i\in C\), set
      $      m_i=\lceil x_i\rceil.$ Return \(m\).
\end{enumerate}

The secondary selection in Step~8 does not require sorting the tie set.
For example, when \(t<1/2\), one may determine the \(k\)-th largest
value among
\[
\{a_i:i\in T\}
\]
using the same deterministic linear-time selection algorithm, include all
indices whose integer parts are strictly larger than this value, and then
include sufficiently many indices at equality to obtain exactly \(k\)
indices. The case \(t>1/2\) is treated analogously using the \(k\)-th
smallest integer part.

\begin{proposition}[Deterministic linear-time complexity]
\label{prop:linear-time-oric}
The Linear-ORIC algorithm returns a solution to Problem~1 for every
\(q\geq 1\). With the secondary selection rule in Step~8, it also returns
a solution to Problem~\ref{problem:4}. The algorithm has deterministic worst-case time
complexity \(O(N)\) and uses \(O(N)\) additional space.
\end{proposition}

\begin{proof}
If \(I=0\), then
\[
\sum_{i=1}^N z_i=0.
\]
Since \(z_i\geq0\), this implies \(z_i=0\) for every \(i\), so
\(x\in\mathbb N^N\). Step~3 returns \(m=x\), which is the unique
solution of Problem~1 and hence also the unique solution of Problem~4.
We may therefore assume below that \(I\geq1\).
Let
$
t=z_{(I)}.$
Every index \(i\in H\) satisfies \(z_i>t\), and hence its fractional part
is strictly larger than the \(I\)-th largest fractional part. Consequently,
every such index belongs to the set of the \(I\) largest fractional parts
and must be rounded upward in every solution characterized by
Proposition~\ref{prop:largest-fractional-parts}.

Similarly, if \(i\in L\), then \(z_i<t\), so \(i\) cannot belong to a set
of \(I\) indices having the largest fractional parts. Such a coordinate
must therefore be rounded downward.
The remaining freedom concerns only the cutoff tie set
\[
T=\{i:z_i=t\}.
\]
Since $
|H|<I\leq |H|+|T|,$
exactly
$
k=I-|H|$
indices from \(T\) must be rounded upward. Hence the vector returned by
the algorithm rounds upward exactly \(I\) coordinates having the \(I\)
largest fractional parts. Proposition~\ref{prop:largest-fractional-parts} therefore implies that the
returned vector solves Problem~1 for every \(q\geq 1\).

If the relative-error tie-breaking rule in Step~8 is used, then among
the indices in \(T\) the algorithm selects the \(k\) largest values of
\(x_i\) when \(t<1/2\), the \(k\) smallest values of \(x_i\) when
\(t>1/2\), and an arbitrary set when \(t=1/2\). Since all indices in
\(T\) have the same fractional part, ordering by \(x_i\) is equivalent
to ordering by \(a_i=\lfloor x_i\rfloor\). Corollary~\ref{corr.tiebreak} therefore implies
that the returned vector also solves Problem~\ref{problem:4}.

It remains to establish the complexity bound. Computing \(a_i\), \(z_i\),
and \(I\) requires \(O(N)\) operations. The cutoff
$ t=z_{(I)} $
can be found in deterministic \(O(N)\) time by a linear-time selection
algorithm. Forming \(H\), \(T\), and \(L\) requires one additional pass
through the coordinates and therefore \(O(N)\) time.

If a secondary tie-break is required, the relevant order statistic inside
\(T\) can be found in \(O(|T|)\) time, and the selected subset can be
constructed in a further \(O(|T|)\) pass. Since \(|T|\leq N\), this step
also requires \(O(N)\) time. All remaining assignments require at most
\(N\) operations. Thus the total deterministic worst-case running time is
$ O(N).$
Storing the fractional parts and the index sets requires \(O(N)\)
additional space.
\end{proof}
\begin{remark}{\em
The complexity statements are made in the real-RAM comparison model,
in which arithmetic operations and exact comparisons of the input
coordinates have unit cost. For floating-point data, equality at the
cutoff should be handled using the precision conventions appropriate
to the application.}
\end{remark}
We note further that Linear-ORIC is optimal for large problems:
\begin{corollary}[Asymptotic optimality]
In the standard RAM model, the worst-case running time
\(\Theta(N)\) of Linear-ORIC is asymptotically optimal.
\end{corollary}
\begin{proof} Returning the full output vector
$
m=(m_1,\ldots,m_N)$
requires writing \(N\) coordinates and therefore takes
\(\Omega(N)\) time. Proposition~\ref{prop:linear-time-oric} gives a matching \(O(N)\)
upper bound. Hence the worst-case running time is \(\Theta(N)\).
\end{proof}

\begin{remark}
If only a solution to Problem~\ref{problem:1} is required, the secondary selection in
Step~8 is unnecessary: any \(k\) indices from the cutoff tie set \(T\)
may be rounded upward. The relative-error ordering is required only when
one also wishes to solve Problem~4.
\end{remark}

\section{Comparison with other rounding methods}\label{sec:comparison}

The sorting-based algorithm  described in
Section~4.1 has \(O(N\log N)\) complexity. The deterministic
selection-based implementation of Section~\ref{subsec:linear-time-oric} avoids full sorting
and has \(O(N)\) worst-case complexity.
The identification of the \(I\) largest fractional
parts is essential and any attempt to bypass it with simpler  rounding methods, whether deterministic \cite{williamson2011} or randomized \cite{raghavan1987}, may fail to yield the optimal solution.
Sorting is one convenient way to perform this step.
Using a worst-case \(O(N\log N)\) comparison sort yields a deterministic
\(O(N\log N)\) implementation. Standard randomized QuickSort yields
expected \(O(N\log N)\) running time.

The asymptotic bias of rounding methods has been  studied by Diaconis \& Freedman \cite{diaconis1979} and more recently by Janson \cite{janson2014}. The argument of vanishing asymptotic bias is often used to argue that these methods yield 'unbiased' solutions.
We will argue here that, unlike what is suggested by the asymptotic properties, the finite sample solution is in fact systematically biased, in a way that has significant implications for  the applications considered in Section \ref{sec:applications}.

\subsection{Fractional rounding}

A  rounding algorithm, often used by default, is {\it fractional rounding}, which
 rounds to $\lfloor x\rfloor $ (resp. $\lceil x\rceil $)  if $x_i-\lfloor x_i\rfloor\leq \theta$ (resp. $x_i-\lfloor x_i\rfloor> \theta$) where $0\leq \theta< 1$. $\theta=1/2$ corresponds to mid-point rounding.
Define
\[
U_\theta(x)=\{i:z_i>\theta\},
\qquad
z_i=x_i-\lfloor x_i\rfloor.
\]
Threshold rounding is feasible precisely when
$
|U_\theta(x)|=I.
$
If \(z_{(I)}>z_{(I+1)}\), any threshold satisfying
\[
z_{(I+1)}\leq\theta<z_{(I)}
\]
produces the ORIC solution. If
$z_{(I)}=z_{(I+1)},$
no pure threshold can in general select exactly $I$ coordinates;
an additional tie-breaking rule is required.

So, a fractional rounding rule may fail to yield the optimal solution to Problem \ref{problem:1} for some $x\in\mathbb{R}_+^N$.
Consequently, no fixed pure-threshold rounding rule is feasible and
optimal for every admissible input \(x\).
\subsection{Independent randomized rounding}\label{sec:randomized}
We now compare the algorithm described above with randomized rounding  \cite{raghavan1987}, in which each $x_i$ is rounded up  with probability $p_i=x_i-\lfloor x_i\rfloor$.
Randomized rounding thus yields a solution given by
\begin{equation}
R_i=\lfloor x_i\rfloor+U_i,
\qquad
U_i\sim\operatorname{Bernoulli}(p_i),
\qquad
p_i=x_i-\lfloor x_i\rfloor.
\label{eq:randomized}
\end{equation}
$$\mathbb{P}(U=u)
=\prod_{i=1}^N
p_i^{u_i}(1-p_i)^{1-u_i},
\qquad
u\in\{0,1\}^N.$$
Assume, without loss of generality, that $1> p_1\geq ...\geq p_N\geq 0.$
Let $I = \sum_{i=1}^N (x_i-\lfloor x_i\rfloor)$.
 Then the optimal solution of Problem \ref{problem:1} corresponds to 
\ba m^*=(\underbrace{1,1,...,1}_{I},0,...0)+\lfloor x\rfloor \label{eq:optimal}\ea
The probability of obtaining this optimum by randomized rounding is
$$\mathbb{P}(R=m^*)=\prod_{i=1}^Ip_i\ \prod_{i=I+1}^N(1-p_i)$$
If the solution is unique, the probability that randomized rounding gives an incorrect solution is 
$$\mathbb{P}(R\neq m^*)=1-\prod_{i=1}^Ip_i\ \prod_{i=I+1}^N(1-p_i)$$
 This probability can be  higher than 50\%, as the following example shows.
\begin{example}
Let  $N= 3$ and $x-\lfloor x\rfloor =(0.4, 0.35, 0.25)$.  The optimal  solution is to round up 0.4 and round down the other components to zero.  Using the formula above the probability of finding the optimum is 
$$ 0.4\times 0.65 \times  0.75= 0.195$$
so the probability of an incorrect result is $0.805.$
\end{example}
Since $\sum p_i=I$, when $N$ is large the normalized discrepancy goes to zero in probability by the  law of large numbers.
Indeed,
\[
\mathbb E\left[
\left(
\frac{1}{N}\sum_{i=1}^N(U_i-p_i)
\right)^2
\right]
=
\frac{1}{N^2}\sum_{i=1}^N p_i(1-p_i)
\leq \frac{1}{4N}.
\]
Hence
\[
\frac{1}{N}\left(\sum_{i=1}^N R_i-M\right)
=
\frac{1}{N}\sum_{i=1}^N(U_i-p_i)
\longrightarrow 0
\]
in \(L^2\), and therefore in probability.
However, the probability of the constraint being satisfied may not go to one.
For example if $p_i=1/2$ and $N$ is even then
$$\mathbb{P}(\sum_{i=1}^N U_i=\frac{N}{2})= \binom{N}{\frac{N}{2}}2^{-N}\sim \sqrt{\frac{2}{\pi N}}\mathop{\to}^{N\to\infty} 0.$$

Independent randomized rounding  \eqref{eq:randomized} generally yields a {\it biased} estimator of the
optimal integer vector \(m^*\). Under a strict cutoff, the sign of
\(\mathbb E[R_i]-m_i^*\) depends on whether the coordinate is rounded
upward or downward by ORIC:
\begin{proposition}[Independent randomized rounding]
Let $x\notin \mathbb{Z}^N$ with $\sum_i x_i=M\in \mathbb{Z}$ and $p_i=x_i-\lfloor x_i\rfloor$. Assume without loss of generality 
$$1> p_1=x_1-\lfloor x_1\rfloor\geq ...\geq p_N=x_N-\lfloor x_N\rfloor \geq  0.$$ 
Define 
$$I=\sum_{i=1}^N p_i=M- \sum_{i=1}^N\lfloor x_i\rfloor\in \mathbb{N}$$ Then $1\leq I\leq N-1$.
Let $U_i\sim$Bernoulli($p_i$) be independent variables with $\mathbb{P}(U_i=1)=p_i$ and define the randomized rounding scheme
$ R_i= \lfloor x_i\rfloor+U_i.$ Then
\begin{enumerate}
    \item Randomized rounding  satisfies the integer constraint {\it in expectation}
    $$E\left( \sum_{i=1}^N R_i\right)=M$$
    \item Assume $p_I>p_{I+1}$ so that the optimal rounding is unique. Then $m_i^*=\lfloor x_i\rfloor+1_{i\leq I}$ and
    $$ E(R_i)-m_i^*= -(1-p_i)1_{i\leq I}+ p_i\ 1_{i> I}$$
\item The probability that randomized rounding produces the optimal vector is 
\begin{equation}
    \mathbb{P}(R=m^*)=\prod_{i=1}^Ip_i \prod_{i=I+1}^N(1-p_i)
\end{equation}
\end{enumerate}
\label{prop:randomized}\end{proposition}
The main justification often advanced for  randomized rounding  is that 'it yields an unbiased estimator'. However $R_i$ is an unbiased estimator of $x_i$, not the desired  integer solution $m_i^*$.

As Proposition \ref{prop:randomized} shows,
 randomized rounding preserves integer constraints in expectation only and, in general, neither satisfies the sum constraint almost surely nor returns the optimal integer rounding. 
It also leads to a systematic upward bias in components whose fractional part is small.
\subsection{Dependent randomized rounding}
Dependent randomized rounding  addresses an important limitation of
independent randomized rounding by introducing dependence between the
coordinatewise rounding decisions. 
cardinality-preserving dependent-rounding scheme of Gandhi et al. \cite{Gandhi2006} is particularly appealing as it enables to preserve the sum constraint.
Applied to the fractional parts
$$
z_i=x_i-\lfloor x_i\rfloor,
\qquad
\sum_{i=1}^N z_i=I\in\mathbb{N},
$$
the cardinality-preserving dependent-rounding scheme of Gandhi et
al.~\cite{Gandhi2006} generates binary random
variables $U_1,\ldots,U_N$ satisfying
$$
\mathbb{P}(U_i=1)=z_i,
\quad i=1,\ldots,N,\quad{\rm and}\qquad
\sum_{i=1}^N U_i=I
\qquad\text{with probability one}.
$$
The rounded vector
$
R_i=\lfloor x_i\rfloor+U_i
$
therefore satisfies
$$
\sum_{i=1}^N R_i=M
\qquad\text{with probability one},
$$
while preserving the coordinatewise marginals.
The resulting binary variables also satisfy negative-correlation properties
which are useful for obtaining concentration bounds for other linear combinations
of the rounded coordinates.

Dependent rounding therefore improves substantially on independent randomized
rounding with respect to feasibility: the sum constraint is satisfied exactly,
rather than only in expectation. It nevertheless addresses a different
criterion from Problem~\ref{problem:1}. 

Independent and dependent randomized rounding have the same expected separable $L^
q$ loss when they use the same coordinatewise marginals, although dependent rounding satisfies the sum constraint almost surely.
But preservation of the marginals does not imply that
each realized integer vector minimizes the rounding error.
Indeed, for every $q\geq 1$,
\begin{align*}
\mathbb{E}\left[
\sum_{i=1}^N |x_i-R_i|^q
\right]
&=
\sum_{i=1}^N
\left[
(1-z_i)z_i^q+z_i(1-z_i)^q
\right].
\end{align*}
This expression depends only on the marginal probabilities $z_i$, and not
on the dependence structure among the variables $U_i$. Since every
realization of dependent rounding is a feasible floor-or-ceiling rounding,
Proposition~3.2 gives
$$
\sum_{i=1}^N |x_i-m_i^*|^q
\leq
\sum_{i=1}^N |x_i-R_i|^q
\qquad\text{with probability one},
$$
where $m^*$ is any ORIC solution. Consequently,
$$
\sum_{i=1}^N |x_i-m_i^*|^q
\leq
\mathbb{E}\left[
\sum_{i=1}^N |x_i-R_i|^q
\right].
$$
When the ORIC solution is unique and $I\geq1$, this inequality is strict:
marginal preservation requires each selected coordinate (i) to be rounded
downward with probability $1-z_i>0$, so dependent rounding returns a
nonoptimal feasible vector with positive probability.

For example, consider the fractional parts
$$
(z_1,z_2,z_3)=(0.4,0.35,0.25),
\qquad I=1.
$$
Any cardinality-preserving dependent rounding with these marginals rounds
upward exactly one coordinate, choosing coordinates $(1,2,3)$ with
probabilities $(0.4,0.35,0.25)$, respectively. The ORIC solution rounds the
first coordinate upward and is therefore obtained with probability $0.4$.
The three possible total absolute errors are, respectively,
$$
1.2,\qquad 1.3,\qquad 1.5,
$$
so dependent rounding has expected total absolute error
$$
0.4 \times 1.2 +0.35 \times 1.3+0.25 \times 1.5=1.31,
$$
whereas ORIC attains the minimum value (1.2) deterministically.

Thus, dependent rounding and ORIC provide complementary guarantees.
Dependent rounding is appropriate when exact feasibility must be combined
with prescribed marginal probabilities, repeated-allocation fairness, or
probabilistic concentration properties. ORIC is appropriate when the objective
is to compute, in a single realization, an exactly feasible integer vector
having minimum rounding error.


\begin{thebibliography}{10}
\bibitem{balinski}
{\sc M.~Balinski and H.~P. Young}, {\em Fair representation: Meeting the ideal
  of One man, One Vote}, Yale University Press, 1982.
  \bibitem{BertholdRENS}
{\sc T.~Berthold},
{\em RENS: The optimal rounding},
\emph{Mathematical Programming Computation},
vol.~6, pp.~33--54, 2014.
\bibitem{BFPRT73}
{\sc M.~Blum, R.~W. Floyd, V.~Pratt, R.~L. Rivest, and R.~E. Tarjan},
{\em Time bounds for selection},
\emph{Journal of Computer and System Sciences},
vol.~7, no.~4,~448--461, 1973.
\bibitem{CoxErnst1982}
{\sc L.~H. Cox and L.~R. Ernst}, {\em Controlled rounding}, INFOR, 20
  (1982), pp.~423--432.
  \bibitem{diaconis1979}
{\sc P.~Diaconis and D.~Freedman}, {\em On rounding percentages}, Journal of
  the American Statistical Association, 74 (1979), pp.~359--364.

\bibitem{FedergruenGroenevelt1986}
{\sc A.~Federgruen and H.~Groenevelt}, {\em The greedy procedure for
  resource allocation problems: Necessary and sufficient conditions for
  optimality}, Operations Research, 34 (1986), pp.~909--918.


\bibitem{fredman2014}
{\sc M.~L. Fredman}, {\em An intuitive and simple bounding argument for
  {Q}uicksort}, Information Processing Letters, 114 (2014), pp.~137 -- 139.
\bibitem{Gandhi2006}
{\sc R.~Gandhi, S.~Khuller, S.~Parthasarathy, and A.~Srinivasan},
  {\em Dependent rounding and its applications to approximation algorithms},
  Journal of the ACM, 53 (2006), pp.~324--360.

\bibitem{grimmett2012}
{\sc G.~R. Grimmett}, {\em European apportionment via the Cambridge
  compromise}, Mathematical Social Sciences, 63 (2012), pp.~68 -- 73.

\bibitem{grotschel1993}
{\sc M.~Gr\"otschel, L.~Lovasz, and A.~Schrijver}, {\em Geometric Algorithms
  and Combinatorial Optimization}, Springer, 1993.

\bibitem{hoare62}
{\sc C.~A.~R. Hoare}, {\em {Quicksort}}, The Computer Journal, 5 (1962),
  pp.~10--16.
\bibitem{HochbaumShanthikumar1990}
{\sc D.~S. Hochbaum and J.~G. Shanthikumar}, {\em Convex separable
  optimization is not much harder than linear optimization}, Journal of the
  ACM, 37 (1990), pp.~843--862.

\bibitem{IbarakiKatoh1988}
{\sc T.~Ibaraki and N.~Katoh}, {\em Resource allocation problems:
  Algorithmic approaches}, MIT Press, Cambridge, MA, 1988.

\bibitem{janson2014}
{\sc S.~Janson}, {\em Asymptotic bias of some election methods}, Annals of
  Operations Research, 215 (2014), pp.~89--136.

\bibitem{korte2002}
{\sc B.~Korte and J.~Vygen}, {\em Combinatorial optimization}, Springer, 2002.


\bibitem{Knuth1995}
{\sc D.~E. Knuth}, {\em Two-way rounding}, SIAM Journal on Discrete
  Mathematics, 8 (1995), pp.~281--290.
  
\bibitem{lenstra1990}
{\sc J.~K. Lenstra, D.~B. Shmoys, and E.~Tardos}, {\em Approximation algorithms
  for scheduling unrelated parallel machines}, Mathematical Programming, 46
  (1990), pp.~259--271.
\bibitem{MarshallOlkinArnold}
    {\sc A.~W. Marshall, I.~Olkin, and B.~C. Arnold},
{\em Inequalities: Theory of Majorization and Its Applications},
2nd ed., Springer, 2011.
\bibitem{niemeyer2008}
{\sc H.~F. Niemeyer and A.~C. Niemeyer}, {\em Apportionment methods}, Math.
  Social Sci., 56 (2008), pp.~240--253.
\bibitem{polya1919b}
{\sc G.~P{\'o}lya}, {\em Proportionalwahl und {W}ahrscheinlichkeitsrechnung},
  Zeitschrift f{\"u}r die gesamte {S}taatswissenschaft/{J}ournal of
  Institutional and Theoretical Economics,  (1919), pp.~297--322.

\bibitem{polya1919}
{\sc G.~P{\'o}lya},  {\em Sur la
  repr{\'e}sentation proportionnelle en mati{\`e}re {\'e}lectorale}, L'{E}nseignement {M}ath{\'e}matique, 20 (1919), ~355--379.

\bibitem{potts1985}
{\sc C.~Potts}, {\em Analysis of a linear programming heuristic for scheduling
  unrelated parallel machines}, Discrete Appl. Math., 10 (1985),~155--164.
\bibitem{Pukelsheim2014}
{\sc F.~Pukelsheim}, {\em Proportional representation: Apportionment methods
  and their applications}, Springer, Cham, 2014.

\bibitem{raghavan1987}
{\sc P.~Raghavan and C.~D. Thompson}, {\em Randomized rounding: a technique for
  provably good algorithms and algorithmic proofs}, Combinatorica, 7 (1987),
  ~365--374.

\bibitem{schulz2002}
{\sc A.~S. Schulz and M.~Skutella}, {\em Scheduling unrelated machines by
  randomized rounding}, {SIAM} Journal on Discrete Mathematics, 15 (2002),
  pp.~450--469.
\bibitem{williamson2011}
{\sc D.~P. Williamson and D.~B. Shmoys}, {\em The design of approximation
  algorithms}, Cambridge University Press, 2011.
  
\end{thebibliography}
\end{document}